\documentclass[10pt]{article}
\usepackage{graphicx}
\usepackage{hyperref}

\def\Title#1{\begin{center} {\Large #1 } \end{center}}
\def\Author#1{\begin{center}{ \sc #1} \end{center}}
\def\Address#1{\begin{center}{ \it #1} \end{center}}

\newcommand\pubblock{\rightline{\begin{tabular}{l} Proceedings of the Fifth Annual LHCP\\ \pubnumber\\
         \pubdate  \end{tabular}}}

\newenvironment{Abstract}{\begin{quotation} \begin{center} 
             \large ABSTRACT \end{center}\bigskip 
      \begin{center}\begin{large}}{\end{large}\end{center} \end{quotation}}

\newenvironment{Presented}{\begin{quotation} \begin{center} 
             PRESENTED AT\end{center}\bigskip 
      \begin{center}\begin{large}}{\end{large}\end{center} \end{quotation}}

\def\Acknowledgements{\bigskip  \bigskip \begin{center} \begin{large}
             \bf ACKNOWLEDGEMENTS \end{large}\end{center}}

\def\beq{\begin{equation}}
\def\eeq#1{\label{#1}\end{equation}}
\def\eeqn{\end{equation}}

\def\beqa{\begin{eqnarray}}
\def\eeqa#1{\label{#1}\end{eqnarray}}
\def\eeqan{\end{eqnarray}}

\let\bar=\overbar

\def\Dslash{\not{\hbox{\kern-4pt $D$}}}
\def\dslash{\not{\hbox{\kern-2pt $\del$}}}

\def\msb{{\bar{\ssstyle M \kern -1pt S}}}

\usepackage{color}

\def\t{{\bar t}}
\def\Mtt{m(t\bar t)}

\def\PTavt{p_{T,{\rm avt}}}
\def\Yavt{y_{\rm avt}}
\def\Ytt{y(t\bar t)}
\def\GeV{\, \rm GeV}

\newcommand\pubnumber{TUM-HEP-1098/17}

\newcommand\pubdate{}

\def\affiliation{
Technische Universit\"{a}t M\"{u}nchen,\\
James-Franck-Str. 1, D-85748 Garching, Germany}

\begin{document}

\large
\begin{titlepage}
\pubblock

\vfill
\Title{Top-quark pair production at NNLO QCD and NLO EW accuracy}
\vfill

\Author{Davide Pagani}
\Address{\affiliation}
\vfill
\begin{Abstract}
We present phenomenological predictions at NNLO QCD and NLO EW accuracy for $t\t$ distributions at the LHC (8 and 13 TeV). 
We discuss the impact of the electroweak corrections and quantify the theory errors from scale and PDF uncertainties. 
Moreover, we show the relevance of a precise determination of the photon PDF for $t\t$ distributions. 

\end{Abstract}
\vfill

\begin{Presented}
The Fifth Annual Conference\\
 on Large Hadron Collider Physics \\
Shanghai Jiao Tong University, Shanghai, China\\ 
May 15-20, 2017
\end{Presented}
\vfill
\end{titlepage}
\def\thefootnote{\fnsymbol{footnote}}
\setcounter{footnote}{0}
%

\normalsize 

%

\section{Introduction}
In these proceeding we present predictions for $t\t$ distributions for the LHC at 8 and 13 TeV, at NNLO QCD accuracy and including also EW corrections. Results are based on the calculation that is described in detail in ref.~\cite{Czakon:2017wor}. We provide results for the following distributions: the top-quark pair invariant mass $m(t\t)$, the average transverse momentum ($\PTavt$) and rapidity ($\Yavt$) of the top and antitop quarks,  and the rapidity $y(t\t)$ of the $t\t$ system. In the case of $\PTavt$ ($\Yavt$) distributions, we average the results for the transverse momentum (rapidity) of the top and the antitop at the histogram level.

We use the same input parameters and choice of scale of ref.~\cite{Czakon:2017wor}, which has been studied and motivated in ref.~\cite{Czakon:2016dgf}. Scale uncertainties are evaluated via the 7-point variation of $\mu_r$ and $\mu_f$ in the standard interval $\{\mu/2,2\mu\}$ with $1/2\leq\mu_r/\mu_f\leq2$. QCD and EW corrections are independently combined for each value of $\mu_{f,r}$.
NNLO QCD predictions are calculated following ref.~\cite{Czakon:2016dgf}, while EW corrections have been obtained in a completely automated way thanks to an extension of the code {\sc\small MadGraph5\_aMC@NLO}  \cite{Alwall:2014hca} that has already been validated in refs.~\cite{Frixione:2015zaa, Badger:2016bpw, Frederix:2016ost}. 

In Section \ref{sec:pheno} we provide phenomenological predictions for the LHC at the 8 and 13 TeV at NNLO QCD and NLO EW accuracy. In fact, we do not include only NNLO QCD and NLO EW corrections but also all the subleading LO and NLO contributions, {\it i.e.} all the terms of $\mathcal{O}(\alpha_s^i \alpha^j)$ with $i+j=2$ and $i+j=3$ as well as the $\mathcal{O}(\alpha_s^4)$ terms. Moreover, we combine QCD and EW corrections in the multiplicative approach, which we denote as ``${\rm QCD \times EW}$'' and represents our best prediction. Details can be found in ref.~\cite{Czakon:2017wor}. We use as PDF set for our best prediction {\sc\small LUXQED} \cite{Manohar:2016nzj,Manohar:2017eqh}, which is NNLO QCD and NLO QED accurate and includes a photon density with a very small uncertainty.

In Section \ref{sec:photonpdf} we show the impact of the photon PDF by comparing predictions based on {\sc\small LUXQED} and {\sc\small NNPDF3.0QED} PDF set \cite{Bertone:2016ume}, similarly to what has been done also in ref.~\cite{Pagani:2016caq}. At variance with {\sc\small LUXQED}, the photon PDF in  {\sc\small NNPDF3.0QED}  gives a contribution with both a large central value and especially uncertainty, showing the relevance of the photon PDF for top distributions.

\section{Phenomenological predictions for the LHC at 8 and 13 TeV}\label{sec:pheno}
\begin{figure}[!h]
\centering
\includegraphics[width=0.4\textwidth]{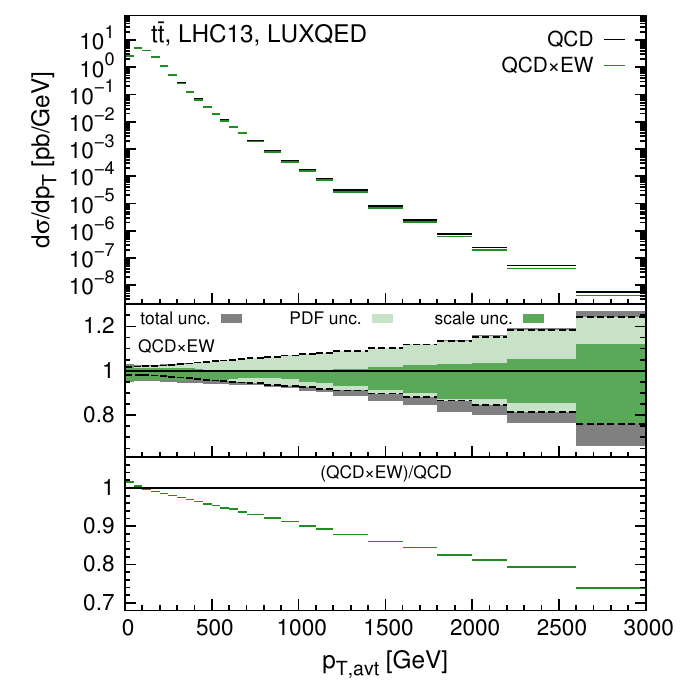}
\includegraphics[width=0.4\textwidth]{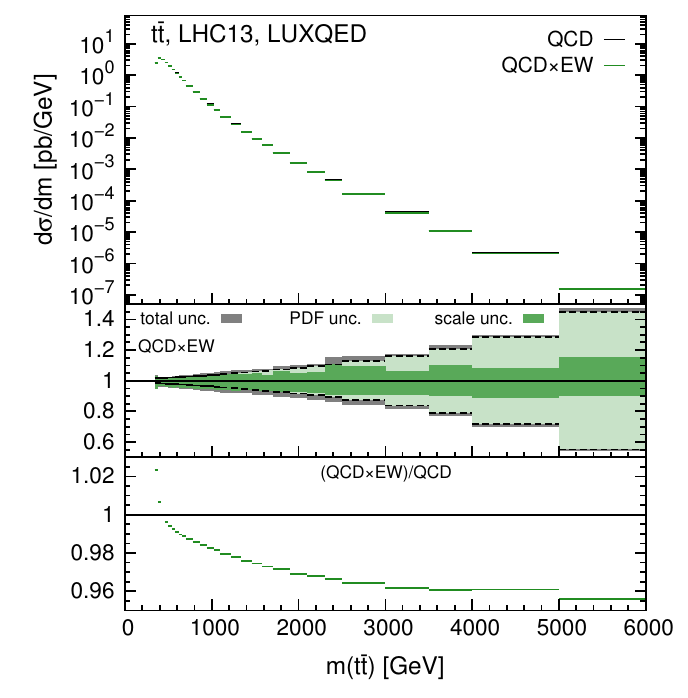}
\includegraphics[width=0.4\textwidth]{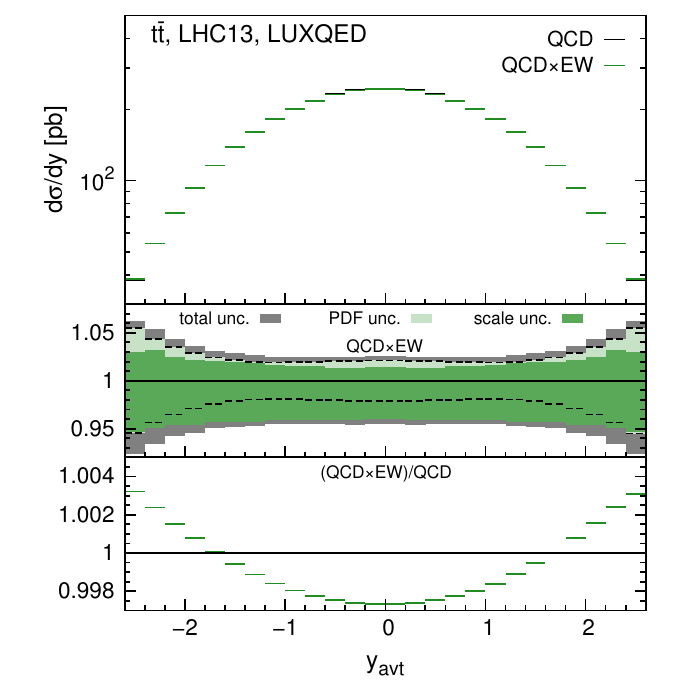}
\includegraphics[width=0.4\textwidth]{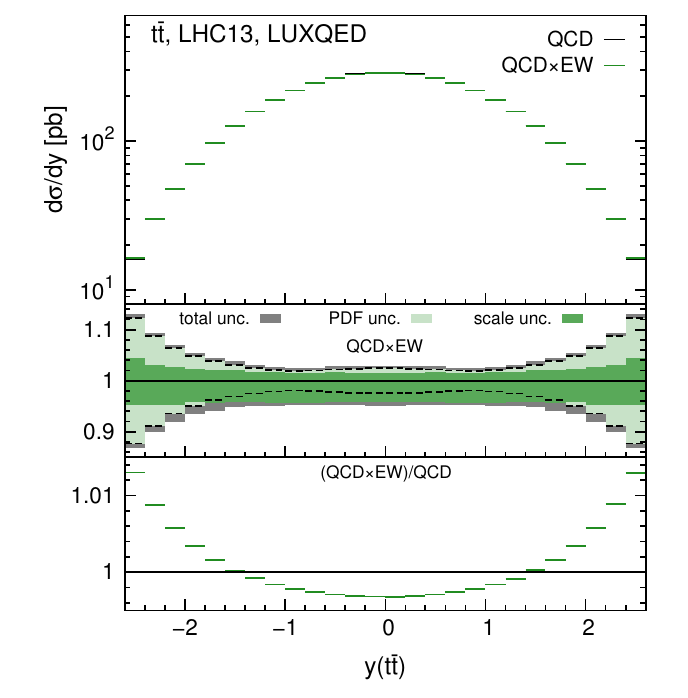}
\caption{Best predictions for LHC 13 TeV. Scale, PDF and  uncertainties are combined in quadrature for each  ${\rm QCD\times EW}$ distribution. Black dashed lines denote the boundaries of the PDF-variation band.  The lower panels shows the ratio of the predictions at ${\rm QCD\times EW}$ and QCD accuracies.}
\label{fig:pheno}
\end{figure}
\begin{figure}[!h]
\centering
\includegraphics[width=0.4\textwidth]{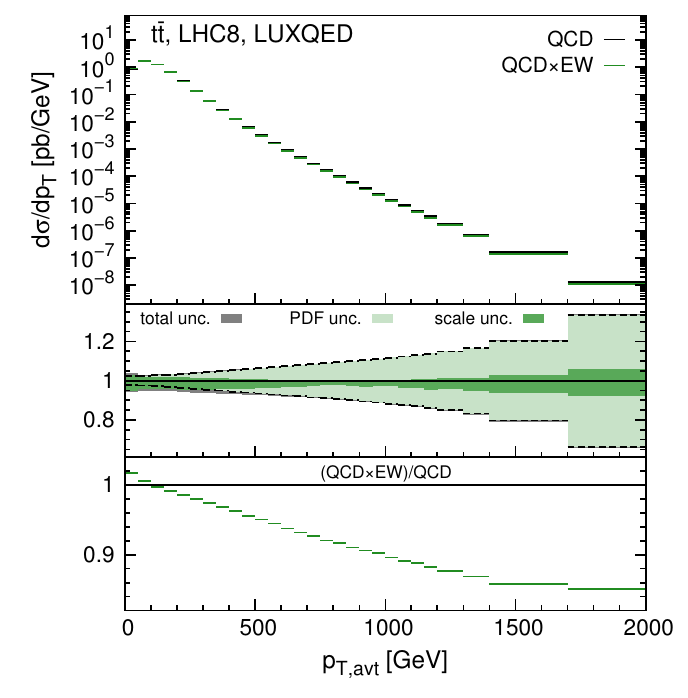}
\includegraphics[width=0.4\textwidth]{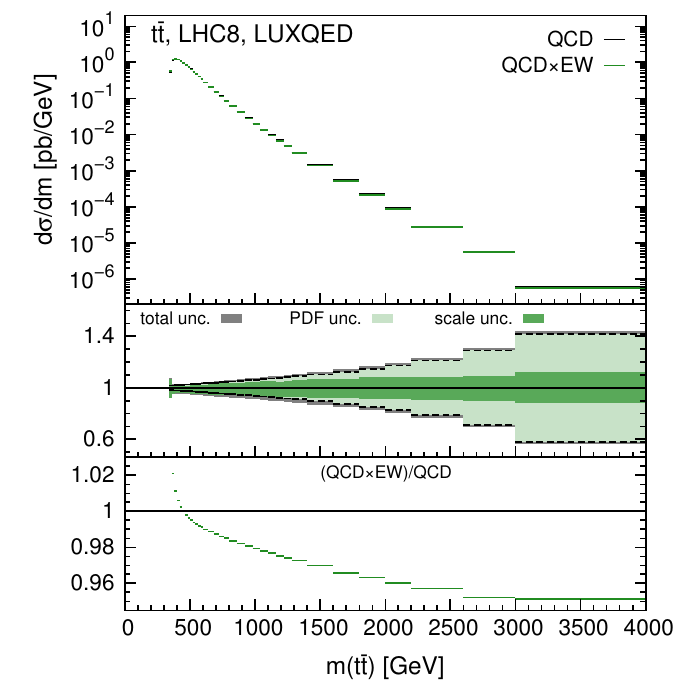}
\includegraphics[width=0.4\textwidth]{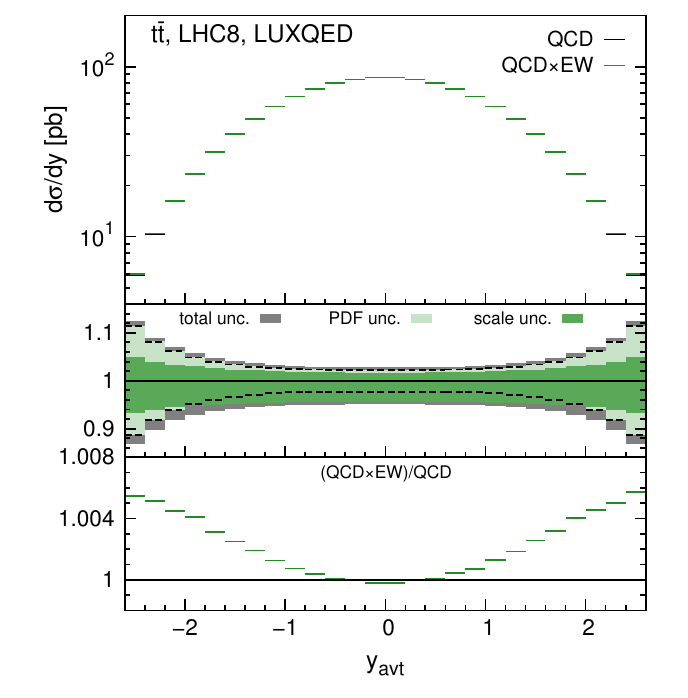}
\includegraphics[width=0.4\textwidth]{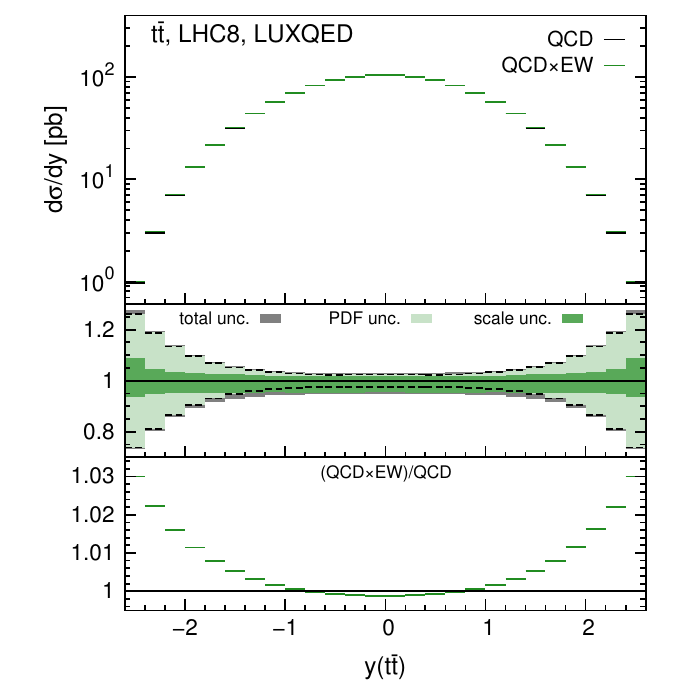}
\caption{Best predictions for LHC 8 TeV. Scale, PDF and  uncertainties are combined in quadrature for each  ${\rm QCD\times EW}$ distribution. Black dashed lines denote the boundaries of the PDF-variation band.  The lower panels shows the ratio of the predictions at ${\rm QCD\times EW}$ and QCD accuracies.}
\label{fig:pheno8}
\end{figure}
Distributions for $m(t\t)$, $\PTavt$, $\Yavt$ and $y(t\t)$ for 13 and 8 TeV are displayed in Figures \ref{fig:pheno} and \ref{fig:pheno8}, respectively. We show our best prediction ${\rm QCD\times EW}$ and the comparison with the result at NNLO QCD accuracy, denoted as QCD. In the first inset we provide the relative  scale and PDF uncertainty as well as their sum in quadrature for the ${\rm QCD\times EW}$ prediction. In the lower inset we show the ratio of ${\rm QCD\times EW}$/QCD, {\it i.e.}, the impact of EW corrections on top on NNLO QCD prediction. Plots in Figure \ref{fig:pheno} (13 TeV) have been taken directly from ref.~\cite{Czakon:2017wor},  while plots in Figure \ref{fig:pheno8} (8 TeV) are available at the web repository \cite{web-based-results}, where many other results can be found. The data files used for drawing plots can also be found as ancillary documentation in the {\sc arXiv} submission for ref.~\cite{Czakon:2017wor}. 

One can see that the impact of  EW corrections  strongly depends on the kinematic distribution, but is in general within the scale(+PDF) uncertainty. The largest corrections  can be observed in the  $\PTavt$ distribution, where they are significant and comparable to the scale-variation band already at $\PTavt\sim 500\GeV$.
Also, the fraction of the theory uncertainty originating from the PDFs strongly depends on kinematics. For the $\Yavt$ and $\Ytt$ distributions, the PDF error is similar to the scale uncertainty for central rapidities, but is larger in the peripheral region, especially for 8 TeV and for the $\Ytt$ distribution.
For large values of $\PTavt$ and $\Mtt$, PDF error are the dominant uncertainty.

By comparing 8 and 13 TeV results we can see that the ${\rm QCD\times EW}$/QCD ratio is very similar for $\PTavt$ and $\Mtt$ and it is slightly enhanced in the peripheral  region in the $\Yavt$ and $\Ytt$ distributions at 8 TeV, similar to the case of the PDF uncertainties. On the other hand, given a value of $\PTavt$ ($\Mtt$) scale uncertainties are typically larger at 13 TeV.

We remind that  these predictions have been obtained in the so-called multiplicative approach for combining QCD and EW corrections. Predictions based on the standard additive approach have been provided in ref.~\cite{Czakon:2017wor}, where we show that, in general, the difference between these two approaches is tiny and well within the scale-uncertainty band. The difference between the two approaches is more pronounced only for large values $\PTavt$  distributions. This kinematic regime is the one where the multiplicative approach is expected to be superior to the additive one and has to be preferred. The multiplicative approach leads to a reduction of  the scale uncertainty, which  in the case of the $\PTavt$ distribution does not overlap with the one from NNLO QCD predictions, however, these features may be sensitive to the choice of the factorisation and renormalisation scales. See ref.~\cite{Czakon:2017wor} for details.

\section{Impact of the photon PDF}\label{sec:photonpdf}
\begin{figure}[!h]
\centering
\includegraphics[width=0.49\textwidth]{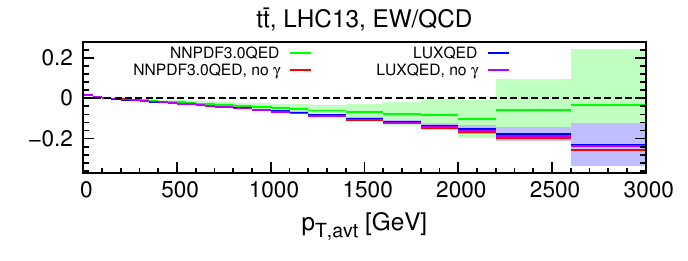}
\includegraphics[width=0.49\textwidth]{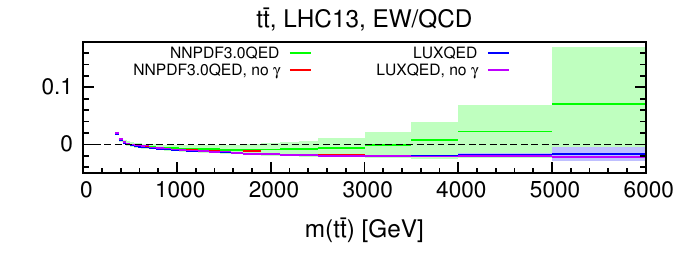}
\includegraphics[width=0.49\textwidth]{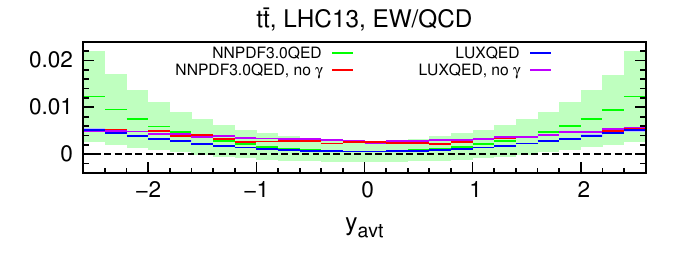}
\includegraphics[width=0.49\textwidth]{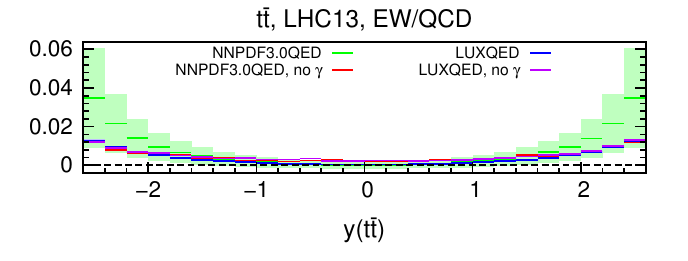}
\caption{Relative impact of EW corrections with and without photon PDF at 13 TeV with {\sc\small LUXQED} and  {\sc\small NNPDF3.0QED} sets.}
\label{fig:photonPDF}
\end{figure}

\begin{figure}[!h]
\centering
\includegraphics[width=0.49\textwidth]{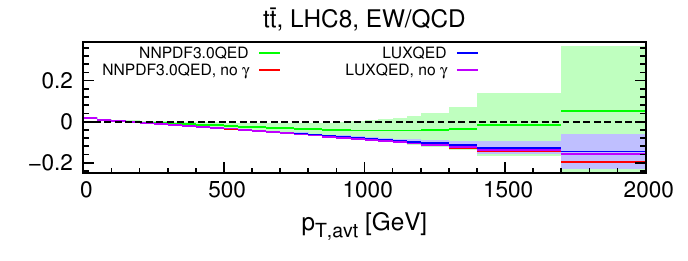}
\includegraphics[width=0.49\textwidth]{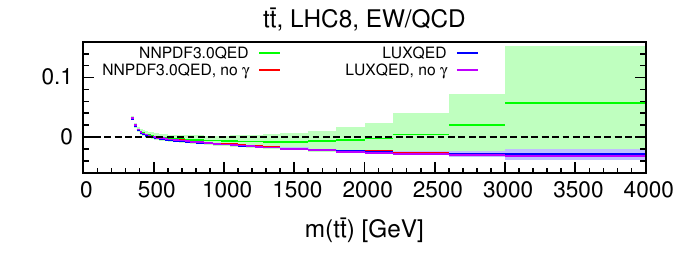}
\includegraphics[width=0.49\textwidth]{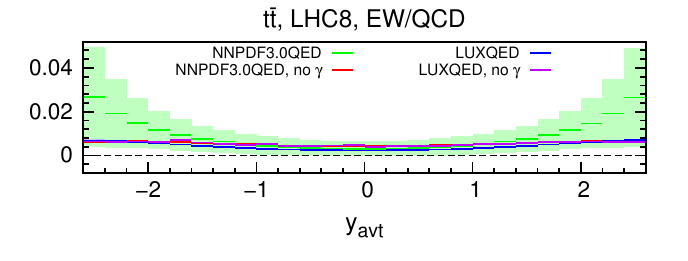}
\includegraphics[width=0.49\textwidth]{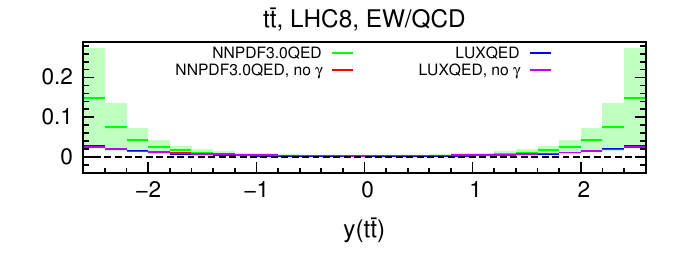}
\caption{Relative impact of EW corrections with and without photon PDF at 8 TeV with {\sc\small LUXQED} and  {\sc\small NNPDF3.0QED} sets.}
\label{fig:photonPDF8}
\end{figure}

In this section we quantify the different impact of the {\sc\small LUXQED} and  {\sc\small NNPDF3.0QED} photon densities on $t\t$ differential distributions at 8 and 13 TeV.  A similar and more detailed comparison has been performed also in ref.~\cite{Pagani:2016caq}, where other PDF sets have been considered. We compare the impact of EW corrections including or not the contribution of the photon PDF. Figure \ref{fig:photonPDF} refers to 13 TeV and it is taken from ref.~\cite{Czakon:2017wor}, while Figure \ref{fig:photonPDF8} refers to 8 TeV.  We shows results for the  for $m(t\t)$, $\PTavt$, $\Yavt$ and $y(t\t)$ distributions and in each plot both  {\sc\small LUXQED} and  {\sc\small NNPDF3.0QED} results, with and without photon PDF, are present.

In each plot we display the EW/QCD ratio, namely the EW corrections divided by the NNLO QCD result. We also show the PDF uncertainty band from the EW corrections only in the cases where the photon PDF is included.

By comparing the difference between the green ({\sc\small NNPDF3.0QED}) and red ({\sc\small NNPDF3.0QED} with the photon PDF set to zero) lines, one can evaluate the impact of the photon PDF in  {\sc\small NNPDF3.0QED}. Similarly, this can be done by comparing the blue ({\sc\small LUXQED}) and violet ({\sc\small LUXQED} with the photon PDF set to zero) lines in the case of {\sc\small LUXQED}.

 As can be seen in Figures~\ref{fig:photonPDF} and \ref{fig:photonPDF8}, while the impact of the photon PDF  is negligible in the case of {\sc\small LUXQED},  with {\sc\small NNPDF3.0QED} it is large for all the $\PTavt$,  $\Mtt$, $\Yavt$ and $\Ytt$ distributions and especially has large uncertainties. The sizable uncertainties that are present at very large $\PTavt$ and  $\Mtt$ for {\sc\small LUXQED} are not induced by the photon but from the other PDFs of the coloured partons, which are probed at large Bjorken-$x$. By setting the photon PDF to zero we have verified that the same argument applies also to  {\sc\small NNPDF3.0QED}.
 
The comparison between plots at 8 and 13 TeV shows that the only case that is sensitive to the energy of the collider is {\sc\small NNPDF3.0QED} with the photon PDF. Not only the central value but also the PDF uncertainties are typically larger in the 8-TeV case. On the contrary, the remaining three cases ({\sc\small LUXQED} with and without the photon PDF and {\sc\small NNPDF3.0QED} with the photon PDF) are all very close and very slightly affected by the energy of the collider.

\section{Conclusions}
We presented results at ${\rm QCD \times EW}$ accuracy for differential distributions with stable top quarks at 8 and 13 TeV. In general, we find that the effect of EW corrections is  within the current total theory uncertainty from PDF and scale variation. However,  in the boosted regime (large $\PTavt$) EW corrections are comparable to the theory error both at 8 and 13 TeV.  All results are available in electronic form \cite{web-based-results} and can be exploited for further studies.

We have also shown that a precise determination of the photon PDF is essential in order to obtain accurate predictions for differential distributions with stable top quarks.

\Acknowledgements
I would like to thank my coauthors of ref.~\cite{Czakon:2017wor}, on which this proceeding is based:  Micha\l{}  Czakon, David Heymes, Alexander Mitov,  Ioannis Tsinikos and Marco Zaro.
The work of D.P is supported by the Alexander von Humboldt Foundation, in the framework of the Sofja Kovalevskaja Award Project ``Event Simulation for the Large Hadron Collider at High Precision''.

\end{document}